\documentclass[aps,twocolumn]{revtex4} 
\usepackage{amssymb,amsmath,amscd}
\usepackage{graphicx} 
\usepackage{bm}
\def\Rpar{{\mathbb R}_{\parallel}}
\def\Rper{{\mathbb R}_{\perp}}
\def\yL{y_{\!_L}} 
\def\yR{y_{\!_R}}

\begin{document}
\setcounter{page}{1}  
\title[]{  
Growth of a One Dimensional Quasiperiodic  Covering with
Locally Determined Decorations.    
}
\author{Hyeong-Chai \surname{Jeong}}  
\email{hcj@sejong.ac.kr}  
\thanks{Fax: +82-2-461-9356}
\affiliation{
Department  of Physics,  Sejong University,  Seoul 143-747,  Korea, 
and \\ 
Asia Pacific Center for  Theoretical Physics, 
Pohang University of Science and Technology,
Pohang 790-784,
Korea
} 

\begin{abstract}
A growth mechanism for a perfect one-dimensional (1D) quasiperiodic 
structure is presented with a local covering rule. 
We use rectangular tiles with two different types of  
string decorations. The string position in a tile
is allowed to move when the tile is attached to an existing patch.
By adjusting the position properly with local information, 
we show that a growth of perfect quasiperiodic structure
is possible. This observation may provide new insight into
how quasicrystals grow with perfect quasiperiodic order.
\end{abstract}



\maketitle


How can quasicrystals grow with quasiperiodic order?
This has been a fundamental question in the quasicrystal community
from the discovery of quasicrystals~\cite{Shechtman84}.
Most quasicrystalline phases are observed in metallic alloys
whose atomic interactions are believed to be short-ranged
while the quasiperiodic arrangement seemingly requires 
non-local information.

The debates on the possibility of a local growth algorithm for a perfect 
quasicrystalline structure were fueled in a quite early stage of quasicrystal 
studies~\cite{Onoda88,Jaric89,Penrose89I,Dworkin95,Grimm02I}. 
The discovery of thermodynamically stable quasicrystalline 
phases~\cite{Tsai87,Conrad98,Tsai00} showed the existence of genuine
quasicrystals, but the deepest question on the physical origin of the
non-crystallographic order has not been fully solved yet.
The various viewpoints on the explanation fall into 
two paradigms: energy-driven perfect quasiperiodic quasicrystals 
and entropy-driven random-tiling quasicrystals.
This energy-entropy debate has mainly focused on the origin of 
stability, but the same arguments can be applied to the growth of
quasicrystals. 
Hence, we can imagine two alternative scenarios for quasicrystal
growth; matching-rule based, energy-driven
growth~\cite{Onoda88} and finite-temperature entropy-driven
growth~\cite{Grimm02I}. A major criticism for the former scenario is that 
non-local information is likely needed for a perfect quasiperiodic
structure growth.  
Such criticisms were mainly based on Penrose's observation 
that no local growth rules can produce a perfect quasicrystalline
tiling in two-dimensional (2D) space~\cite{Penrose02B,Dworkin95}. 
Recently, we provided a local
growth algorithm to produce a perfect quasicrystalline structure in 3D
and showed that non-local information is not necessary for growth with
the perfect quasiperiodic order~\cite{Jeong07prl}.  
However, the algorithm has an artificial element and works only
with a special type of seed. Also, it can produce only one
particular type of Penrose tiling known as the 
cartwheel tiling~\cite{Gardner77}. 

\begin{figure}[t!] 
\includegraphics[width=80mm]{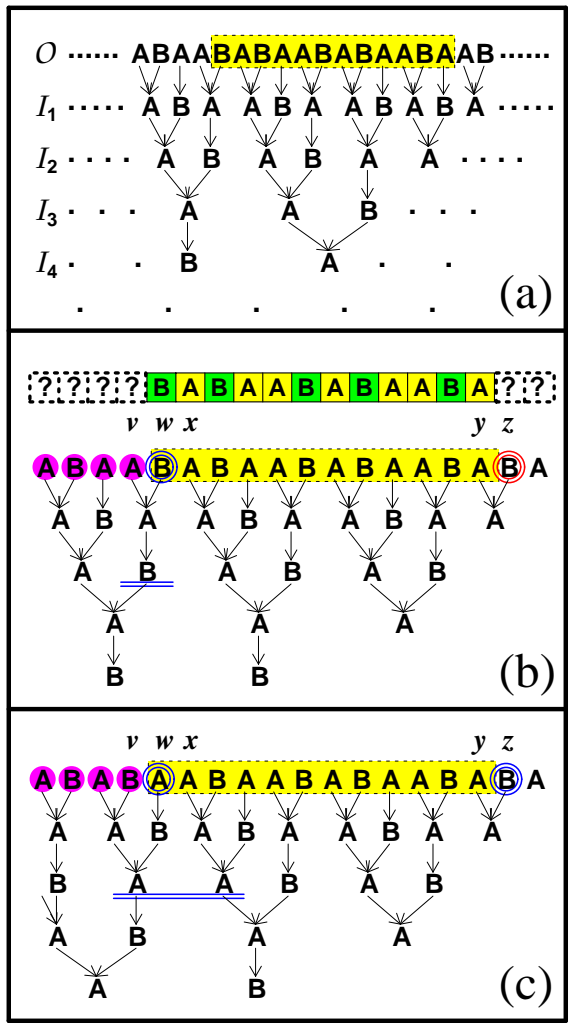} 
\caption[0]{(Color online) 
(a) Fibonacci lattice composed of two types of tiles, 
$A$ and $B$. A Fibonacci lattice is a 1D infinite arrangement 
of two types of tiles, which allows an infinite sequence of
``inflations" ($A B \rightarrow A$ and $A \rightarrow B$).
(b) Growth from a finite patch of a Fibonacci lattice.
To determine the type of attaching tile, sometimes one needs
to investigate the whole configuration of the existing patch. 
The types of tiles at sites $w$ and $z$ are correlated,
as are those at sites $w$ and $v$. The tile at site $z$
should be $A$ when $B$ is at $w$. Otherwise, the patch becomes a
forbidden $BB$ segment after 4 inflations. 
(c) If the tile at $w$ is $A$, then attaching a $B$-tile at site
$z$ is correct because it produces a legal $AB$ segment after 4
inflations. For the growth (purple circled tiles) in the left side of 
the patch, see the text and the end note~\cite{loc.type}.
}
\label{f.fibo}
\end{figure}

In this letter, we introduce a new type of local growth algorithm
and show that perfect quasiperiodic structures can be grown with a local
process even in 1D systems. Our growth rule is local in the sense that it
determines the type of attaching tile according to the
decoration (information) of the tile to which the tile is attached.
However, it is different from conventional growth rules in a couple of
aspects. First, we use the covering and allow overlaps between
neighboring tiles. Second, the position of decoration is not fixed
prior to the attachment. We choose the decoration position 
when we attach the tile to the existing patch of tiles. 
The latter may not be a typical way of decoration
for a matching rule in conventional tilings but seems to be
physically plausible for the covering process. 
Physically, a tile usually represents a stable cluster atoms, 
but the positions of atoms can be adjusted a bit when it overlaps
with neighboring clusters. 

One can easily see that non-local information is needed for 
the growth of a 1D perfect quasiperiodic structure
with a conventional growth rule in which decorations for the matching 
rule, as well as the shapes of the tiles, are fixed.
Figure~\ref{f.fibo}(a) shows a Fibonacci lattice,
one of the most well known 1D quasiperiodic system and its ``inflations''. 
A Fibonacci lattice is a special 1D infinite arrangement of two types
of tiles, say, $A$ and $B$, which allows an infinite sequence of
inflations, a composition $A B \rightarrow A$ and $A \rightarrow
B$ shown in Fig.~\ref{f.fibo}. 
To allow an inflation, a Fibonacci lattice should
be decomposed as an infinite array of $AB$ and $A$ segments. 
In other words, there should not be $BB$ segments in any part of a
Fibonacci lattice. Furthermore, the inflation of a
Fibonacci lattice should be another (infinite) Fibonacci lattice 
because it allows infinite iterative inflations. 
This requirement excludes $AAA$ segments,
which would produce $BB$ segments after an inflation. By the same
token, an $ABABAB$ segment is not allowed because its inflation 
would produce an $AAA$ segment whose (next) inflation becomes a $BB$
segment.  The longer segments must be investigated to exclude 
the ``forbidden" $BB$ segments in the more inflated lattices. Hence, 
in order to grow a Fibonacci lattice, we need 
a non-local growth rule, which provides information on an increasing
range of tile arrangement as the patch grows, as illustrated in 
Figs.~\ref{f.fibo}(b) and~(c).

Consider a growth from a ``correct" (subset of a
Fibonacci lattice) finite patch shown in the upper part of
Fig.~\ref{f.fibo}(b). 
When a tile is attaching to site $v$, its type is determined
by one or two nearest sites. If a $B$-tile is at site $w$, 
an $A$-tile should be at site $v$ to avoid the forbidden
configuration $BB$.  If an $A$-tile is at site $w$ instead, as
shown in Fig.~\ref{f.fibo}(c), a $B$-tile should be 
at site $v$ to avoid an $AAA$ segment, which produces the forbidden
$BB$ segment after one inflation~\cite{loc.type}. 
Therefore, the tile type at site $v$ can be locally determined. 
However, there are cases in which one needs to know the whole configuration of
the existing patch to determine the type of the attaching tile.  
When we attach a tile to site $z$, we can determine its
type only after investigating the whole lattice because the tile type at
site $z$ is related to the tile type at site $w$. 
An $A$-tile at site $w$ [Fig.~\ref{f.fibo}(a) and~(b)]
enforces a $B$-tile at site $z$  
while a $B$-tile at site $w$ [Fig.~\ref{f.fibo}(c)]
enforces a $A$-tile at site $z$.
If both tiles at sites $w$ and $z$ are $B$ type,
as shown in Fig.~\ref{f.fibo}(b), then the patch
produces the forbidden $BB$ segment after 4 inflations. 
Similarly, placing $A$-tiles at both sites $w$ and $z$ 
results in the forbidden $BB$ segment after 5 inflations.  
If site $w$ were empty when we attach a tile to site $z$, 
both $A$ and $B$-tiles would be possible at site $z$. 
In any case, to place a correct type of tile at site $z$, 
we need to know information about site $x$, whether it is $A$, 
$B$ or empty.

The example of Fig.~\ref{f.fibo}(b) illustrates why non-local 
information is needed for the Fibonacci tiling growth. 
Although the arrangement of 13 tiles from $w$ to $z$ 
is incorrect (not a subset of Fibonacci lattice), any connected proper
subset of the arrangement is correct. In other words, both 
sub-configurations (of $w$ to $y$ and of $x$ to $z$) 
with 12 consecutive tiles can be found in a Fibonacci lattice. 
Therefore, a ``deception" of length 13~\cite{loc.length} 
is possible even if we use a rule that checks for the arrangement of
11 tiles (and allow only correct configurations of length 12).
Here, deception means a legal (satisfying the growth rule), but
incorrect, configuration~\cite{Dworkin95}.
Since the growth process does not allow tiles to be removed, a
deception (which is not a part of a Fibonacci tiling) cannot grow to a
Fibonacci tiling, so we need a growth rule that allows no 
deceptions of any size. With this requirement, we
can show that a Fibonacci tiling cannot be grown 
with conventional local growth rules, where tile shapes are
fixed and a finite range of tile configurations are investigated 
when we attach a tile. 
Let us start with a range~1 growth rule~\cite{loc.length}; 
attach an $A$ or $B$-tile to an (existing) $A$-tile, but attach only 
an $A$-tile to a $B$-tile. This simple growth rule forbids forming 
a $BB$ arrangement and guarantees that length~2~\cite{loc.length}
configurations are correct. 
However, it allows a length~3 deception, an $AAA$ arrangement.
We can avoid three tile deceptions by introducing a more restricted
growth rule that allows only correct three tile configurations.
However, the new 
growth rule can make a deception on a larger scale, for example, a
three tile deception of inflated tiles such as an $ABABAB$ arrangement.
Since a deception can be made in all scales of multiply inflated tile
sizes~\cite{Gardner77}, it is unavoidable for a local growth rule
with conventional tiles. 

\begin{figure}[t!] 
\includegraphics[width=80mm]{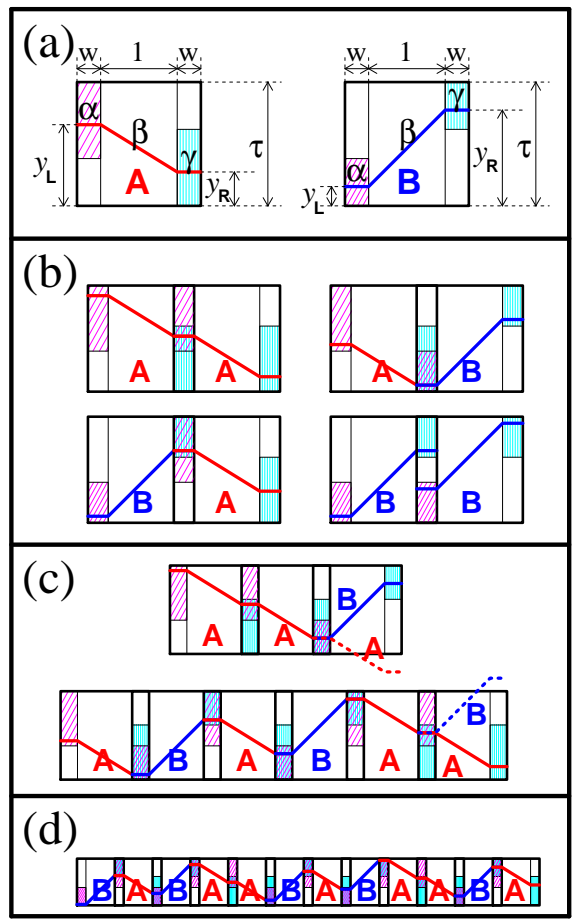} 
\caption[0]{(Color online) 
(a) Basic building units for the covering
with two types of decorations, deco-$A$ and deco-$B$.
Each tile is a $(1+2w) \times \tau$ rectangle, and its 
decoration consists of three line segments, 
$\alpha$, $\beta$, and $\gamma$. 
The slopes of the $\beta$-segments depend 
on the decoration types and are given by $\beta_A = -1/\tau$ and
$\beta_B = 1$ 
for $A$ and $B$ types respectively.  
The vertical position of the string decoration is movable in the
bounded regions. The vertical coordinate $\yL$ of the $\alpha$
segments should satisfy $\yL \in [1/\tau,\tau)$ for deco-$A$
and $\yL \in [0,1/\tau)$ for deco-$B$ so that the segment $\gamma$
remains in the tile ($\yR = \yL-1/\tau \ge 0$ for deco-$A$
and $y_B^R = \yL+1 < \tau$ for deco-$B$).
(b) Overlap rule for the Fibonacci covering. 
Two neighboring rectangles in a covering overlap
in the $W$-region. A rectangle can be overlapped
with its neighbor only if the string of the 
rectangle can adjust its vertical position 
so that its string decoration coincides with that of
the neighboring rectangle in the $W$-region.
This excludes a $BB$ arrangement and permits three pairwise
overlaps, $AA$, $AB$, and $BA$ arrangement, for the
isolated two-tile configurations. 
(c) Growth of a Fibonacci covering. The overlap rule excludes
incorrect configurations, such as $AAA$ or $ABABAB$ 
arrangement (denoted by dotted lines), by allowing only 
a $B$-tile to an $AA$ segment and an $A$-tile to an $ABABA$ segment.
(d) Fibonacci covering corresponding to Fig.~\ref{f.fibo}(a) with
string decorations. String decorations can remain inside the
rectangles when they are arranged as a Fibonacci lattice.  
}
\label{f.cover}
\end{figure}

In this paper, we consider the covering with locally adjustable
decorations and show that a perfect 1D quasiperiodic 
structure can be grown by using a local growth rule.
We use rectangular tiles with two 
different string decorations shown in  Fig.~\ref{f.cover}(a). 
The string position is not fixed until
the tile is attached to the existing patch of tiles. 
The lattice structure obtained by our growth is the Fibonacci 
lattice considered in Fig.~\ref{f.fibo}(a). Covering is distinguished from 
tiling because it allows overlaps between basic building
units. However, in 1D structures, the relationship between coverings
and tilings is trivial. 
We can always construct a covering corresponding to a given tiling by 
enlarging the basic building units. The mapping can be reversed to
produce a tiling from a covering, reducing the basic building units
until the units join edge-to-edge (point-to-point in pure 1D system)
without overlap. Although the lattice structures of the covering and
the tiling are equivalent, we use covering here because our adjustable
decoration is more naturally expressed in terms of covering. 
 
Figure~\ref{f.cover} shows the basic building units and 
our growth rule. We use a rectangle tile with the size of 
$(1+2w)\times \tau$ as the basic building unit.
Here, $\tau = (1+\sqrt{5})/2$ is the golden mean, and $w>0$ is
an arbitrary positive 
number. 
In the figure, we choose $w=\frac{1}{2\tau}$ so that the unit
becomes a square. There are two types of decorations, deco-$A$ and 
deco-$B$, as shown in Fig.~\ref{f.cover}(a). 
Both decorations consist of three line segments, $\alpha$, $\beta$,
and $\gamma$. The segments $\alpha$ and $\gamma$ are horizontal
lines with lengths of $w$ for both decorations, but 
the segments $\beta$ have different slopes.  
For deco-$A$, the slope is $\beta_A = -1/\tau$ 
while for deco-$B$, it is $\beta_B = 1$. 
Their vertical positions are movable in the bounded regions denoted
by the hatched lines of purple and skyblue. 
The vertical coordinate $\yL$ of the segment 
$\alpha$ should satisfy 
$\yL \in \mathbb{D}_A^\alpha = [1/\tau,\tau)$ for deco-$A$
while that of deco-$B$ is given by 
$\yL \in \mathbb{D}_B^\alpha = [0,1/\tau)$ 
so that their $\gamma$-segments remain in the tiles 
($\yR = \yL-1/\tau \ge 0$ for deco-$A$
and $\yR = \yL+1 < \tau$ for deco-$B$).
In the growth process, a tile is added to an existing patch
only if the string of the tile can adjust its vertical position 
so that its string decoration coincides with that of
the existing patch in the overlapped region. 
This excludes $BB$ arrangement and permits three ways of pairwise
overlaps, $AA$, $AB$, and $BA$ arrangements, for the
isolated two tile configurations, as shown in Fig.~\ref{f.cover}(b).  
Both an $A$-tile (tile with deco~$A$) and 
a $B$-tile (tile with deco~$B$) 
can be added to an $A$-tile~\cite{loc.addtoA},
but only an $A$-tile can be added to a $B$-tile 
because the $\gamma$-region
of deco-$B$ does not overlap with the $\alpha$-region of deco-$B$.
As the patch grows, the position of the string decoration 
at the boundary effectively carries the information 
on all tiles in the patch and forces 
the correct type of tile to be added.
Figure~\ref{f.cover}(c) illustrates how our overlap rule excludes
an incorrect configuration such as $AAA$ or $ABABAB$.
For an $AA$ arrangement, $\yR$ of the right $A$-tile is always
smaller than $\tau-2/\tau = 1/\tau^2$ while $\yL$ of an $A$-tile 
cannot be smaller than $1/\tau$. If we attach an
$A$-tile and move the string to the required position, it goes out of
the tile, as shown by the dotted line in Fig.~\ref{f.cover}(c). 
When one type of tile is excluded, the other type of
tile is always allowed because the $\alpha$-regions of an $A$-tile and 
those of a $B$-tile are complementary to each other 
($\mathbb{D}_A^\alpha \cup \mathbb{D}_B^\alpha = [1,\tau)$
and $\mathbb{D}_A^\alpha \cap \mathbb{D}_B^\alpha = \emptyset$).
Therefore, a $B$-tile is forced to attach to an $AA$ arrangement. 
Similar argument shows that an $A$-tile is forced to an $ABABA$
arrangement. As illustrated in the lower panel of Fig.~\ref{f.cover}(c),
the $\gamma$-segment position of the rightmost $A$-tile $\yR$ cannot
be smaller than $2-2/\tau=2/\tau^2$, which is larger than the
maximum $\yL$ value of a $B$-tile, $1/\tau$.

We have illustrated how our overlap rules exclude the incorrect
arrangement of tiles. Now, we see that our string decoration is 
possible for any correct configuration of Fibonacci lattice.
In Fig.~\ref{f.cover}(d), string decorations are shown for
the Fibonacci covering corresponding to Fig.~\ref{f.fibo}(a).
They coincide in all overlapped regions and remain in the
unit cells.

\begin{figure}[t!] 
\includegraphics[width=80mm]{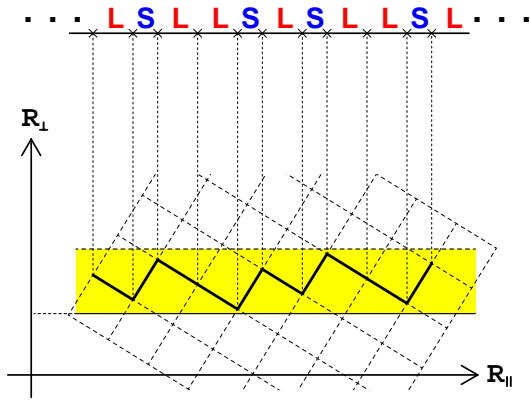} 
\caption[0]{(Color online) 
The lifting of a (pure 1D) Fibonacci lattice into a 2D hyper-space.
A Fibonacci tiling consist of two types of tiles, say
$L$ and $S$, can be lifted into a 2D square lattice
whose $x$-axis has the slope $-1/\tau$ with respect to the 1D tiling
on $\Rpar$ by mapping $L$ to a $x$-edge and $S$ to a $y$-edge of 
a square. The lengths of tiles $L$ and $S$ are given by $\cos\theta$ and
$\sin\theta$, respectively, where $\tan \theta = 1/\tau$.
The embedded step (solid thick line) in the 2D lattice can
be covered by a strip parallel to $\Rpar$ with width $\Delta =
\cos\theta + \sin\theta$ when the position of the strip is well
chosen. 
The $\Rper$ coordinates of the step decrease by $\sin\theta$ for an
$L$-tile and increase by $\cos\theta$ for an $S$-tile.
}
\label{f.lift}
\end{figure}

We can show that Fibonacci lattices are grown by our local rule
by lifting the covering to a 2D hyperspace. 
Any 1D tiling consisting of two types of tiles, 
say an $L$-tile and an $S$-tile, can be lifted into a ``representative 
surface", a step (denoted by solid thick line in Fig.~\ref{f.lift}) on
a 2D square lattice by mapping one type to an $x$-edge and the other type to
a $y$-edge of a square in the 2D space. 
Let us first consider a pure 1D Fibonacci tiling with (pure 1D) tiles
whose lengths are given by $\cos\theta$ and $\sin\theta$ for tiles $L$
and $S$, respectively, where $\theta = \arctan(1/\tau)$, as shown in
Fig.~\ref{f.lift}.  
Then, we see that the $\Rper$ coordinates of a representative surface (the
step) in the hyper-space decreases by $\sin\theta$ for an
$L$-tile and increases by $\cos\theta$ for an $S$-tile.
Using this, we can calculate the $\Rper$ coordinate changes for
all series of inflated tiles and show that the embedded step in the 2D
hyper-space can be covered by a strip parallel to $\Rpar$ with width
$\Delta = \cos\theta + \sin\theta$ when the position of the strip is well
chosen~\cite{Katz86,Jeong01JPA}.  
We can also show the converse, a 1D tiling whose embedded step  
is covered by the strip (with the $\Rper$ width of $\Delta$), 
is a Fibonacci tiling, by considering the mapping of the strip under
inflation~\cite{Katz86,Goodman89B}.  
Now, we need to show that the string decoration of our covering can be 
mapped to the embedded step in the hyper space and that our growth 
rule, indeed, forces the step to be in a strip of proper width. 
Note that the overall length scale in the $\Rper$ space is irrelevant
as long as the ratio of the $\Rper$ space coordinates of the two tile, 
$r =  \Delta \Rper(L)/\Delta \Rper(S) = - 1/\tau$, is fixed. If
both $\Delta \Rper(L)$ and $\Delta \Rper(S)$ are increased by a factor
of $\lambda$, then we can still get the Fibonacci lattice by simply 
increasing the strip width by the same factor. 
For numerical simplicity, we choose $\Delta \Rper = |\yR-\yL|$ 
of $A$-tile and $B$-tiles (which corresponds to $L$ and $S$, respectively)
be $-1/\tau$ and $1$, respectively, satisfying $r = - 1/\tau$. 
Hence, $A$-tiles and $B$-tiles would be
arranged as a Fibonacci lattice when we choose the strip width 
as $1+1/\tau = \tau$. We have considered a 1D arrangement 
of rectangle tiles (instead of pure 1D tiles of $L$ and $S$)
to encode the information of the embedded step into
the string decoration. Now, the role of the strip is replaced by
the height of the rectangles and, hence, is given by $\tau$. 
The horizontal length of the tiles can have (fixed) finite, but
any positive, values because we are interested in the sequence of 
$A$ and $B$ types, not the actual positions of vertices. 
The lengths in the pure 1D tiles of Fig.~\ref{f.lift},
$\cos\theta$ and $\sin\theta$, are chosen for the lifting to have a
simple geometrical interpretation.

Although the mapping between the string decoration of our covering
and the embedded step of Fibonacci tiling is simple, 
the physical implication of our growth algorithm is rather 
surprising. Contrary to conventional wisdom, it shows that
a perfect 1D quasiperiodic structure can be grown with 
a pure local growth process. Debates on the possibility of
a local growth algorithm for a quasiperiodic system has 
focused on 2D or 3D systems with the assumption that the 
local growth of a 1D quasiperiodic structure is impossible.
Here, we have shown that the information on the whole sequence of
$A$ and $B$ types can be encoded into the position of the string
and passed to the attached new tile by a local process. 
In fact, this is the way that the long-range translational order
is created in the crystal growth process. In crystal growth, 
a tile (unit-cell) joins to the existing patch of tiles side-by-side
such that it fills the space without overlaps or gaps.
By doing so, it carries information on the positions of all existing
tiles; from its own position, we can calculate all possible positions
of tiles compatible with its position although we do not know how many
tiles already exist in the patch. 
The same information-transferring process happens
in our quasicrystal growth process. From the position of string,
we do not know how many tiles exist, but we can calculate all
possible positions of the given types of tiles.  

Previous local growth algorithms
for the quasiperiodic structures~\cite{Onoda88,Jeong07prl} are
considered to be too complex for atoms to follow. 
Furthermore, they could produce only certain particular
types of quasiperiodic structures with specially prepared seeds.   
The growth algorithm presented here is believed to apply for 
general quasiperiodic structures. By adjusting the 
decoration positions in the seeds, all classes of Fibonacci lattice
structures~\cite{Jeong01JPA} can be obtained.
Furthermore, by changing the shapes of the decorations and the heights
of the tiles, we can grow all 1D quasiperiodic structures, which 
can be obtained by using the projection methods~\cite{Janot92B}.
This new type of local growth algorithm may help us to answer the
old puzzle of how quasicrystals grow with quasiperiodic order. 
Atomic structures of many quasicrystals, especially those that show 
high-quality quasiperiodic ordering, are well described by 
using quasi-unit-cell
models~\cite{Steinhardt98Nat,Jeong03AC,Jeong03PRB} based on the
covering with overlapping tiles. The overlap corresponds physically to the 
sharing of atoms by neighboring clusters. 
The overlaps are restricted to certain relative positions and 
orientations of the tiles according to overlap rules.  
Hence, it is a physically plausible assumption 
that decorations adjust their positions
when they overlap with existing clusters of atoms. 
An important implication of this work is that atoms may relax 
their positions in such a way that 
they carry long-range information
on the atomic positions in other tiles and allow 
quasiperiodic structure to be grown by a local process. 
We hope that our observation can provide new insights as to how
quasicrystals can grow in a structure with long-range quasiperiodic
ordering.  

\begin{acknowledgments}
We would like thank Korea Institute for Advanced Study (KIAS)
for the hospitality during our visit. 
This work was supported by the Korea Research
Foundation Grant funded by the Korean Government (MOEHRD),
KRF-2006-312-C00172.
\end{acknowledgments}


\begin{thebibliography}{10}
\bibitem{loc.type}
Note that the three tiles (denoted by first three purple circles) at
the left of 
  site $y$ must be in an $ABA$ configuration regardless of the tile
  type at site $y$ to avoid a $BB$ or an $AAA$ configuration in the
  twice inflated lattice, 
  as illustrated by the blue double lines in Figs.~1(b) and~(c).
\bibitem{loc.length}
The length of a tile arrangement [patch, segments, etc.] are defined as the
  number of tiles in the arrangement [patch, segments]. We also define the
  range of a growth rule similarly. It is the number of tiles that are
  investigated to determine the type of the attaching tile.
\bibitem{loc.addtoA}
For $\yR < 1/\tau$, an $A$-tile is forced;
otherwise, a $B$-tile is forced.
\end{thebibliography}

\end{document}